# Survey of Publicly Available State Health Databases

Sean Hooley and Latanya Sweeney

Harvard University
Cambridge, Massachusetts
shooley@fas.harvard.edu, latanya@fas.harvard.edu

## ABSTRACT

We surveyed every state and the District of Columbia to see what patient specific information states release on hospital visits and how much potentially identifiable information is released in those records. Thirty-three states release hospital discharge data in some form, with varying levels of demographic information and hospital stay details such as hospital name, admission and discharge dates, diagnoses, doctors who attended to the patient, payer, and cost of the stay. We compared the level of demographic and other data to federal standards set by the Health Information Portability and Accountability Act or HIPAA), which states do not have to adhere to for this type of data. We found that states varied widely in whether their data was HIPAA equivalent; while 13 were equivalent (or stricter) with demographic fields only 3 of the 33 states that released data did so in a form that was HIPAA equivalent across all fields.

## INTRODUCTION

People expect that the information they tell their doctor will remain private, and that expectation extends to doctors they see at the hospital. Doctor-patient confidentiality makes the relationship work to its fullest - if there is no fear the doctor will discuss private medical issues, a patient can feel secure telling his doctor important details, leading to better care and better data. Some states require hospitals to share information about each patient encounter, and the states in turn, may sell or give the data away. The released version does not include people's names, but does include demographic information about the patient and details about the visit. Most people are unaware these data exist, much less that they are shared publicly. Individuals could be harmed if the data could be matched back to the patient because it contains diagnoses that may include drug and alcohol dependency, tobacco use, venereal diseases, and other sensitive information, even if that was not the reason for the hospitalization. It seems prudent to survey the decisions states make when sharing these data to see how they compare to the federal standard for sharing patient level health information to see if standards are the same.

It is important to understand that sharing data beyond the patient encounter offers many worthy benefits to society. These data may be particularly useful because they contain a complete set of hospital discharges within the state, thereby allowing comparisons across regions and states such as rating hospital and physician performances and assessing variations and trends in care, access, charges and outcomes. Research studies that have used these datasets include: examinations of utilization differences based on proximity [1], patient safety [2,3], and procedures [4]; and, a comparison of motorcycle accident results in states with and without helmet laws [5]. The very completeness that make these studies informative makes it impossible to rely on patients to consent to sharing because the resulting data would not be as complete.

Of course, when data are shared publicly, the information becomes available for many other purposes too, some that may not be as motivating. A recent Bloomberg news article reported that the top multi-state buyers of patient level hospital data are commercial and other for-profit organizations, not researchers [6].

The challenge is to find ways comprehensive patient level data can be shared widely so society can enjoy the benefits of data sharing without risks of harms to individuals.

## BACKGROUND

When a person goes to the hospital, information about her is recorded and in most states it is passed on to the state government or a separate nonprofit organization that collects that information for the state. Additionally, many states use the Federal-State-Industry partnership Healthcare Cost and Utilization Project (HCUP) to collect the information for them. Then the state, nonprofit or HCUP distributes this information (with names and some geographic and temporal information redacted) to the public. This flow of information is authorized in most states by a legislative mandate. Depending on the state, different levels of information are publicly available at varying costs, and some states require approval to obtain the information. Some even have different tiers of data, with variable restrictions and costs.

The Health Information Portability and Accountability Act (HIPAA) in the United States is the federal regulation that dictates sharing of medical information beyond the immediate care of the patient, prescribing to whom and how physicians, hospitals and insurers may share a patient's medical information broadly. Not all health data is covered by HIPAA, but for medical data covered by HIPAA to be shared publicly, all dates must be in years and only the first 3 digits of the patient's ZIP code (totally omitted, with only the state name if the population in the ZIP code is less than 20,000) can be released.[1] The information states distribute about hospital stays is not covered by HIPAA, so states may make different decisions.

## METHODS

We performed a survey of the information each state makes publicly available as well as the cost and restrictions for the data. This was done by visiting a state's website, using online search engines to search for "inpatient data" or "discharge data" for each state, and utilizing a subscription service to see what information each state released to the public. Some states were also contacted by email or phone.

[1] 45 CFR 164.514(b)(2) (2007).

We began by using The National Association of Health Data Organizations (NAHDO) website, a membership and educational association that maintains a web site with information on 49 states (all but Alabama) and the District of Columbia. Each state has a web page with information about the collection and release policies of their healthcare data as well as links related to that information such as the states health care organization or department, contact people, and, where relevant, the law(s) that mandate the availability of the information. Many states also had information on their websites about obtaining this data. A few, like Vermont, were free but most had at least a nominal fee and several were thousands of dollars (see Table 3).

Given time and monetary restrictions, we did not acquire every state's health data. However, some of the states data we did acquire differed in the information they released from that reported on their website. For instance, Washington State reported on their website that they release age in years, but in fact release age in months as well in a separate data field. Virginia reported releasing 9 digit zip codes, but the data we received showed only the first 5 digits. To populate the tables in the Results section, we used fields with naming that we understood such as "AGE_GROUP" or "ZIPCODE" assisted by data dictionaries the states provide to decode the fields. However, some fields may be reporting information not readily apparent without intimate knowledge of that state's data. The information presented here was culled from many sources and the best effort was made to collect the most accurate and up-to-date information.

Some states have Data Use Agreements that require acknowledging (by clicking on the state's website) or signing forms agreeing to comply with the agreement. There was a great deal of variability in what the agreements required including restrictions on who could use the data and what they could do with it as well as who they could share it with and how long they were allowed to keep and use the data. Since HIPAA does not allow a Data Use Agreement to offset its standards, terms of Data Use Agreements are not considered in this paper.

## RESULTS

We organized our information into four tables. Thirty-three states provide some publicly available hospital information (see map in Figure 1). Nebraska was listed on HCUP as providing data, but NAHDO says they do not, so we considered Nebraska as not sharing hospital data for the purposes of this paper. The information presented here was culled from many sources and the best effort was made to collect the most accurate and up-to-date information. Please send the authors any updates/corrections for rectification. Check dataprivacylab.org/projects/50states for updates.

Table 1 lists the demographic information released by each state including gender, address, and age. All the states that provide data give the patients gender. Address information was the address of the insurance policy holder, which is usually the patient's home. Ten states release 3 digit ZIP codes for addresses, subject to further masking if the ZIP code has a small population. Maine and South Carolina only provide the county name. West Virginia and Nevada provide no address information and Rhode Island stopped providing any in 2007. While all the geographic information would be HIPAA equivalent, Colorado, New York and Washington State provide birth month information, which would not be allowed if the data were covered by HIPAA. Seven states released age in age groups, which is stricter than HIPAA regulations. Age groups were variable in size though most were 5-year groups with different size groupings for children and infants. Missouri released the birth year and the rest of the states' age data were released as age in years. Either would be HIPAA compliant.

Table 2 shows if and how admit date, discharge date and discharge status are released. For example, Virginia releases the year and quarter of admission and discharge as well as length of stay. All the states that released discharge data released discharge status, such as "Routine discharge" or "Dsch/Trnf to skilled nursing facility w/Medicare", or if the patient died at the hospital. Five states released the date in the admission and/or discharge data, and 21 others released the month or quarter along with year. None of this information would be HIPAA compliant; only 7 states released HIPAA conforming year-only dates (hour or day of week are not restricted by HIPAA) for both admission and discharge fields.

Table 3 lists where to get the data, cost, and if the state has a mandate to release the data. Forty states have a legal mandate to collect hospital data (not all distribute it though). Some states like Washington and New Hampshire distribute the data directly, some like Virginia work through separate nonprofits to do so, while 14 (not including Nebraska) rely on HCUP to collect and distribute the information. Several states that distribute information directly or through a nonprofit also have their information available through HCUP, though the data available through HCUP may be a different price and may offer different fields than data directly from the state. Prices ranged from free to ten thousand dollars for a year's amount of data, and often had discounts for educational institution's or other non-profits and had different pricing for data sets with more potentially identifiable information. The costs reported here are to research institutions for the inpatient-unrestricted version of the public data file from the most recent year available.

Table 4 shows whether a state's data would be HIPAA compliant. The hospital data released by states is not covered under HIPAA, but we assessed whether it would be equivalent to HIPAA rules in Table 4. This was created using Tables 1 and 2 and assessing whether the data was equivalent to HIPAA standards - in this case all dates reported year minimally and geographic information was minimally 3 digit ZIP codes. Interestingly, six states' demographic data not only adhered to HIPAA standards, but was stricter. However, for many states, the admission and discharge information they release was not HIPAA equivalent, only one of those states was among the three states whose data would be fully HIPAA compliant.

Figure 2 shows a map of the states whose demographic data would be HIPAA compliant, states whose data release is stricter than HIPAA and the three states that would not be HIPAA equivalent as detailed in Table 1. Figure 3 shows a map of how the 13 states whose demographic data is HIPAA equivalent drops to 3 states when admission and discharge data is screened for HIPAA equivalence.

| State | Gender | Address | Age[1] |
|---|---|---|---|
| Alabama | | | |
| Alaska | | | |
| Arizona | Yes | 5 digit zip code | In Years |
| Arkansas | Yes | 3 digit zip code | In Years |
| California | Yes | 3 digit (or nothing if not unique) subject to masking | In Years (subject to masking) |
| Colorado | Yes | 3 digit zip code | Birth month and year |
| Connecticut | | | |
| Delaware | | | |
| District of Columbia | | | |
| Florida | Yes | 5 digit zip code | In Years |
| Georgia | | | |
| Hawaii | Yes | 5 digit zip code | Age Group (Birth Year in HCUP) |
| Idaho | | | |
| Illinois | Yes | 3 digit zip code | Age Group |
| Indiana | | | |
| Iowa | Yes | 5 digit zip code | In years |
| Kansas | | | |
| Kentucky | Yes | 5 digit zip code | In years |
| Louisiana | | | |
| Maine | Yes | County | In Years |
| Maryland | Yes | 3 digit zip code | In Years |
| Massachusetts | Yes | 3 digit zip code | In Years |

**Table 1a. Comparison of demographic data in patient-specific hospital discharge data by state, Alabama through Massachusetts. Diagonal line pattern indicates that State does not release public data. [1]Age groups are variable in size. Many are groups of 5 years with different size groupings for children and infants.**

| State | Gender | Address | Age[1] |
|---|---|---|---|
| Michigan | | | |
| Minnesota | | | |
| Mississippi | | | |
| Missouri | Yes | First 3 digits if first 3 digits of ZIP has population >20,000 | Birth year |
| Montana | | | |
| Nebraska[2] | | | |
| Nevada | Yes | State | In Years |
| New Hampshire | Yes | 5 digit zip code | In Years |
| New Jersey | Yes | 5 digit zip code | In Years |
| New Mexico | Yes | 3 digit zip code | In Years |
| New York | Yes | 5 digit zip code | Birth month and year |
| North Carolina | Yes | 5 digit zip code | In Years |
| North Dakota | | | |
| Ohio | | | |
| Oklahoma | Yes | 5 digit zip code | Age Group |
| Oregon | Yes | 3 digit zip code | In Years |
| Pennsylvania | Yes | 5 digit zip code | In Years |
| Rhode Island | Yes | Removed in 2007 | In Years |
| South Carolina | Yes | County | Age Group |
| South Dakota | Yes | 5 digit zip code | In Years |
| Tennessee | Yes | County, 5 digit zip code | In Years |
| Texas | Yes | 5 digit zip code (first 3 if a ZIP code has fewer than 30 cases) | Age Group (expanded for HIV/drug/alcohol) |

**Table 1b. Comparison of demographic data in patient-specific hospital discharge data by state, Michigan through Texas. Diagonal line pattern indicates that State does not release public data. [1]Age groups are variable in size. Many are groups of 5 years with different size groupings for children and infants. [2]Nebraska (via NAHDO) says they do not release but HCUP says that they release data.**

| State | Gender | Address | Age[1] |
|---|---|---|---|
| **Utah** | Yes | 5 digit zip code | Age Group |
| **Vermont** | Yes | 3 digit zip code (categories; 5-digit if pop>10k) | Age Group |
| **Virginia** | Yes | 5 digit zip code | In Years |
| **Washington** | Yes | 5 digit zip code | In months |
| **West Virginia** | Yes | State | In years |
| **Wisconsin** | Yes | 5 digit zip code | In years |
| **Wyoming** | | | |

**Table 1c. Comparison of demographic data in patient-specific hospital discharge data by state, Utah through Wyoming. Diagonal line pattern indicates that State does not release public data. [1]Age groups are variable in size. Many are groups of 5 years with different size groupings for children and infants.**

| State | Admission Date | Discharge Date | Discharge Status |
|---|---|---|---|
| Alabama | | | |
| Alaska | | | |
| Arizona | Year, Month, Hour | Year, Month, Hour, Length of Stay | Yes |
| Arkansas | Year, Month, Hour | Year, Month, Hour, Length of Stay | Yes |
| California | Year, Quarter | Year, Length of Stay | Yes |
| Colorado | Year, Month, Date, Hour | Year, Month, Day of Week, Length of Stay | Yes |
| Connecticut | | | |
| Delaware | | | |
| District of Columbia | | | |
| Florida | Year, Hour | Year, Length of Stay | Yes |
| Georgia | | | |
| Hawaii | Year, Month | Year, Month, Length of Stay | Yes |
| Idaho | | | |
| Illinois | Year, Quarter | Year, Quarter | Yes |
| Indiana | | | |
| Iowa | Year, Month, Date | Year, Month, Day of Week | Yes |
| Kansas | | | |
| Kentucky | | Year, Quarter, Length of Stay | Yes |
| Louisiana | | | |
| Maine | Date | Date | Yes |
| Maryland | Year, Month | Year, Length of Stay | Yes |
| Massachusetts | Year, Month | Year, Length of Stay | Yes |

**Table 2a. Comparison of admission and discharge fields in patient-specific hospital discharge data by state, Alabama through Massachusetts. Diagonal line pattern indicates that State does not release public data.**

| State | Admission Date | Discharge Date | Discharge Status |
|---|---|---|---|
| Michigan | | | |
| Minnesota | | | |
| Mississippi | | | |
| Missouri | Year, Hour | Year, Hour, Length of Stay | Yes |
| Montana | | | |
| Nebraska[1] | | | |
| Nevada | Year, Month, Hour | Year, Month, Hour, Length of Stay | Yes |
| New Hampshire | Year, Hour | Year, Hour | Yes |
| New Jersey | Year, Month, Hour | Year, Hour, Length of Stay | Yes |
| New Mexico | Year, Month, Hour | Year, Month, Hour, Length of Stay | Yes |
| New York | Year, Month | Date | Yes |
| North Carolina | Year, Month | Year, Month | Yes |
| North Dakota | | | |
| Ohio | | | |
| Oklahoma | Year, Month, Day of Week | Year, Month, Day of Week | Yes |
| Oregon | | Year, Length of Stay | Yes |
| Pennsylvania | Year, Day of Week, Hour | Year, Day of Week, Hour, Length of Stay | Yes |
| Rhode Island | Year, Month | Year, Month, Length of Stay | Yes |
| South Carolina | Year, Month, Day of Week | Year, Month, Day of Week, Length of Stay | Yes |
| South Dakota | Year, Month | Year, Length of Stay | Yes |
| Tennessee | Date, Hour | Date | Yes |
| Texas | Year, Day of Week | Year, Quarter, Length of Stay | Yes |

**Table 2b. Comparison of admission and discharge fields in patient-specific hospital discharge data by state, Michigan through Texas. Diagonal line pattern indicates that State does not release public data. Nebraska (via NAHDO) says they do not release but HCUP says that they release data.**

| State | Admission Date | Discharge Date | Discharge Status |
|---|---|---|---|
| Utah | | Year, Quarter, Length of Stay | Yes |
| Vermont | | Year, Length of Stay | Yes |
| Virginia | Year, Quarter | Year, Quarter, Length of Stay | Yes |
| Washington | Year, Hour | Month, Hour, Length of Stay | Yes |
| West Virginia | | Year, Length of Stay | Yes |
| Wisconsin | | Year, Quarter, Length of Stay | Yes |
| Wyoming | | | |

**Table 2c. Comparison of admission and discharge fields in patient-specific hospital discharge data by state, Utah through Wyoming. Diagonal line pattern indicates that State does not release public data.**

| State | Mandate | Organization | Through HCUP | Cost (per year unless noted) |
|---|---|---|---|---|
| Alabama | | www.alaha.org | | Not Release |
| Alaska | No | Alaska State Hospital and Nursing Home Association. http://www.ashnha.com/ | | Not Release |
| Arizona | Yes | Arizona Department of Health Services. http://www.azdhs.gov/ | Yes | $135 |
| Arkansas | Yes | Arkansas Department of Health. http://www.healthy.arkansas.gov/ | Yes | $485 |
| California | Yes | California Office of Statewide Health Planning & Development. http://www.oshpd.ca.gov/ | Option[2] | $200 |
| Colorado | No | Colorado Hospital Association. http://www.cha.com/ | | $0.02 for record |
| Connecticut | Yes | Connecticut Office of Health Care Access. http://www.ct.gov/ohca/ | | Not Released |
| Delaware | Yes | Delaware Health Statistics Center, Division of Public Health. http://www.dhss.delaware.gov /dhss/dph/hp/healthstats.html | | Not Released |
| District of Columbia | No | District of Columbia Hospital Association. http://www.dcha.org/ | | Not Released |
| Florida | Yes | Florida Center for Health Information and Policy Analysis. http://ahca.myflorida.com/SCHS/ | Option[2] | $100 / year ($25/quarter) |
| Georgia | Yes | GHA: An Association of Hospitals & Health Systems. http://www.gha.org/ | | Not Released |
| Hawaii | No | Hawaii Health Information Corporation. http://www.hhic.org/ | Yes | $1,035 |
| Idaho | | Idaho Hospital Association. http://www.teamiha.org/ | | Not Released |
| Illinois | Yes | Illinois Department of Public Health. http://www.idph.state.il.us/ | | $100+ |
| Indiana | Yes | Indiana Hospital Association. http://www.ihaconnect.org/ | | Not Released |
| Iowa | Yes | Iowa Hospital Association. http://www.ihaonline.org/ | Yes | $585 |
| Kansas | Yes | Kansas Hospital Association. http://www.kha-net.org/ | | Not Released |
| Kentucky | Yes | Kentucky Cabinet for Health and Family Services-Office of Health Policy. http://www.chfs.ky.gov/ohp/ | Yes | $1,535 |
| Louisiana | Yes | Louisiana Department of Health and Hospitals. http://www.dhh.louisiana.gov/ | | Not Released |
| Maine | Yes | Maine Health Data Organization. http://www.maine.gov/mhdo/ | Option for some years[2] | $675 |
| Maryland | Yes | Health Services Cost Review Commission. http://www.hscrc.state.md.us/ | Yes | $35 |
| Massachusetts | Yes | Division of Health Care Finance and Policy. http://www.mass.gov/dhcfp | Yes | $835 |

**Table 3a. Administrative information by state, Alabama through Massachusetts. [2]Data available through HCUP may be different price and may offer different fields than one from state [7]. [3]Cost to research institutions for inpatient unrestricted version of public data file for most recent year available.**

| State | Mandate | Organization | Through HCUP | Cost (per year unless noted) |
|---|---|---|---|---|
| **Michigan** | No | Michigan Health & Hospital Association. http://www.mha.org/ | | Not Released |
| **Minnesota** | Yes | Minnesota Hospital Association. http://www.mnhospitals.org/ | | Not Released |
| **Mississippi** | Yes | Mississippi Dept of Health, Office of Health Informatics. http://www.msdh.ms.gov/phs | | Not Released |
| **Missouri** | Yes | Missouri Department of Health and Senior Services. http://www.dhss.mo.gov/ | | $150+ |
| **Montana** | No | MHA - An Association of Montana Health Care Providers. http://www.mtha.org/ | | Not Released |
| **Nebraska** | No | Nebraska Hospital Association. http://www.nhanet.org/ | Yes[1] | Not Released |
| **Nevada** | Yes | Healthcare Cost and Utilization Project (HCUP). http://www.hcup-us.ahrq.gov/sidoverview.jsp | | $435 |
| **New Hampshire** | Yes | New Hampshire Department of Health & Human Service. http://www.dhhs.nh.gov/ | | Free |
| **New Jersey** | Yes | New Jersey Department of Health & Senior Services. http://www.state.nj.us/health/ | Yes | $60 |
| **New Mexico** | Yes | New Mexico Department of Health. http://www.health.state.nm.us/ | Yes | $485 (2009 and 2010 available) |
| **New York** | Yes | New York State Dept of Health. http://www.health.state.ny.us/ | Option[2] | ~$350, final price at request time |
| **North Carolina** | Yes | Cecil G. Sheps Center for Health Services Research at the University of North Carolina at Chapel Hill. http://www.shepscenter.unc.edu | Yes | $535 |
| **North Dakota** | Yes | North Dakota Department of Health. http://www.ndhealth.gov/ | | Not Released |
| **Ohio** | No | OHA: Ohio Hospital Association. http://www.ohanet.org/ | | Not Released |
| **Oklahoma** | Yes | Oklahoma State Department of Health. http://www.ok.gov/health/ | | $50-7500 |
| **Oregon** | Yes | Office for Oregon Health Policy and Research. http://www.oregon.gov/OHPPR/ | | $250 |
| **Pennsylvania** | Yes | Pennsylvania Health Care Cost Containment Council (PHC4). http://www.phc4.org/ | | $4,500 |
| **Rhode Island** | Yes | Rhode Island Department of Health. http://www.health.state.ri.us/ | Option[2] | $115 |
| **South Carolina** | Yes | South Carolina State Budget & Control Board, Office of Research and Statistics. http://ors.sc.gov/ | Option[2] | $1.25 per 1,000 records with a minimum $100 |
| **South Dakota** | Yes | South Dakota Association of Healthcare Organizations. http://www.sdaho.org/ | Yes | $785 |
| **Tennessee** | Yes | Tennessee Hospital Association (THA) Health Information Network.[4] http://www.tha-hin.com/ | | $10,000 |
| **Texas** | Yes | Texas Health Care Information Collection, Center for Health Statistics, Texas Department of State Health Services. www.dshs.state.tx.us/thcic/ | | $6,000 recent years, 1999-2006 free |

**Table 3b. Administrative information by state, Michigan through Texas. [1]Nebraska (via NAHDO) does not release but HCUP says they release data. [2]Data available through HCUP may be different price and offer different fields than one from state [7]. [3]Cost to research institutions for inpatient unrestricted public data file for most recent year available. [4] Office of Health Statistics, Tennessee Department of Health also has an order form for a public use file, though the fields released are not listed. http://health.state.tn.us/statistics/index.htm**

| State | Mandate | Organization | Through HCUP | Cost (per year unless noted) |
|---|---|---|---|---|
| **Utah** | Yes | Office of Health Care Statistics, Utah Department of Health. http://www.health.utah.gov/hda/ | Option[2] | $3,150 |
| **Vermont** | Yes | Division of Health Care Administration. http://www.bishca.state.vt.us/ | Option[2] | Free |
| **Virginia** | Yes | Virginia Health Information. http://www.vhi.org/ | | $975-$2500 |
| **Washington** | Yes | Washington State Department of Health. http://www.doh.wa.gov/ | | $50 |
| **West Virginia** | Yes | West Virginia Health Care Authority. http://www.hcawv.org/ | Yes | $510 |
| **Wisconsin** | Yes | Wisconsin Hospital Association. http://www.wha.org/ | Yes | $835 |
| **Wyoming** | No | Wyoming Hospital Association. http://www.wyohospitals.com/ | | Not Released |

**Table 3c. Administrative information by state, Utah through Wyoming. [2]Data available through HCUP may be different price and offer different fields than one from state [7]. [3]Cost to research institutions for inpatient unrestricted public data file for most recent year available.**

| State | HIPAA Equivalence for Demographic Data[1] | HIPAA Equivalence for Admission & Discharge[2] | HIPAA Equivalence or Better for Both[3] |
|---|---|---|---|
| Alabama | | | |
| Alaska | | | |
| Arizona | No | No | |
| Arkansas | Yes | No | |
| California | Yes | No | |
| Colorado | No | No | |
| Connecticut | | | |
| Delaware | | | |
| District of Columbia | | | |
| Florida | No | Yes | |
| Georgia | | | |
| Hawaii | No | No | |
| Idaho | | | |
| Illinois | Stricter | No | |
| Indiana | | | |
| Iowa | No | No | |
| Kansas | | | |
| Kentucky | No | No | |
| Louisiana | | | |
| Maine | Stricter | No | |
| Maryland | Yes | No | |
| Massachusetts | Yes | No | |

**Table 4a. Assessment of HIPAA equivalence by state, Alabama through Massachusetts. Diagonal line pattern indicates that State does not release public data. [1,2] HIPAA equivalent if ZIP is only 3 digits and dates (including age) given only in years. "Stricter" if all fields were HIPAA equivalent and at least one is stricter. "Less strict" if any fields were not equivalent to HIPAA standard. [3]Only "yes" responses reported. All blanks that do not have diagonal pattern are "no".**

| State | HIPAA Equivalence for Demographic Data[1] | HIPAA Equivalence for Admission & Discharge[2] | HIPAA Equivalence or Better for Both[3] |
|---|---|---|---|
| Michigan | | | |
| Minnesota | | | |
| Mississippi | | | |
| Missouri | Yes | Yes | Yes |
| Montana | | | |
| Nebraska[4] | | | |
| Nevada | Stricter | No | |
| New Hampshire | No | Yes | |
| New Jersey | No | No | |
| New Mexico | Yes | No | |
| New York | No | No | |
| North Carolina | No | No | |
| North Dakota | | | |
| Ohio | | | |
| Oklahoma | No | No | |
| Oregon | Yes | Yes | Yes |
| Pennsylvania | No | Yes | |
| Rhode Island | Stricter | No | |
| South Carolina | Stricter | No | |
| South Dakota | No | No | |
| Tennessee | No | No | |
| Texas | No | No | |

**Table 4b. Assessment of HIPAA equivalence by state, Michigan through Texas. Diagonal line pattern indicates that State does not release public data. [1,2] HIPAA equivalent if ZIP is only 3 digits and dates (including age) given only in years. "Stricter" if all fields were HIPAA equivalent and at least one is stricter. "Less strict" if any fields were not equivalent to HIPAA standard. [3]Only "yes" responses reported. All blanks that do not have diagonal pattern are "no". [4]Nebraska (via NAHDO) says they do not release but HCUP says that they release data.**

| State | HIPAA Equivalence for Demographic Data[1] | HIPAA Equivalence for Admission & Discharge[2] | HIPAA Equivalence or Better for Both[3] |
|---|---|---|---|
| Utah | No | No | |
| Vermont | No | Yes | |
| Virginia | No | No | |
| Washington | No | No | |
| West Virginia | Stricter | Yes | Yes |
| Wisconsin | No | No | |
| Wyoming | | | |

**Table 4c. Assessment of HIPAA equivalence by state, Utah through Wyoming. Diagonal line pattern indicates that State does not release public data. [1,2] HIPAA equivalent if ZIP is only 3 digits and dates (including age) given only in years. "Stricter" if all fields were HIPAA equivalent and at least one is stricter. "Less strict" if any fields were not equivalent to HIPAA standard. [3]Only "yes" responses reported. All blanks that do not have diagonal pattern are "no".**

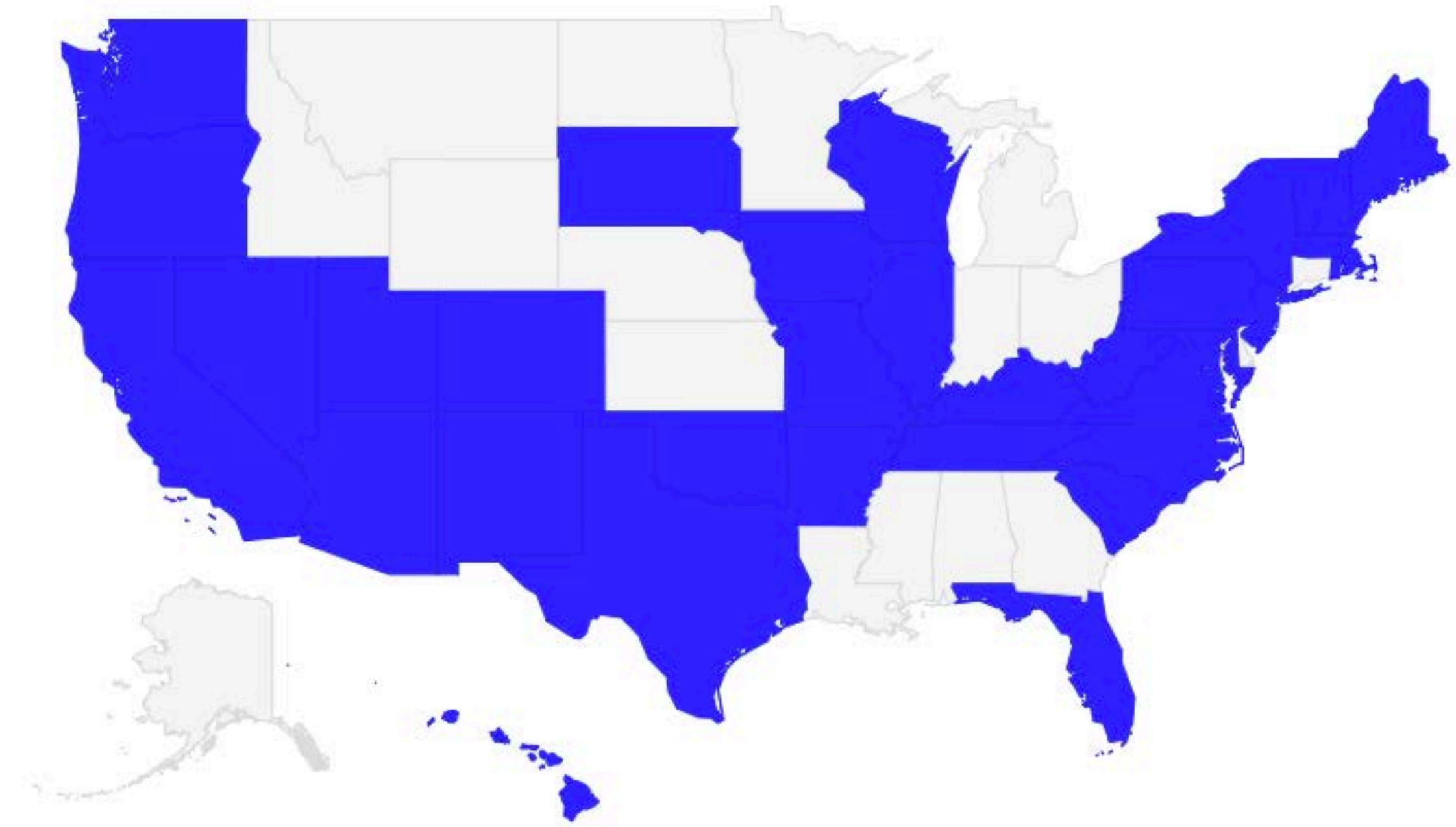

**Figure 1. United States map showing states that release patient-level hospital data in blue, for a total of 33 states.**

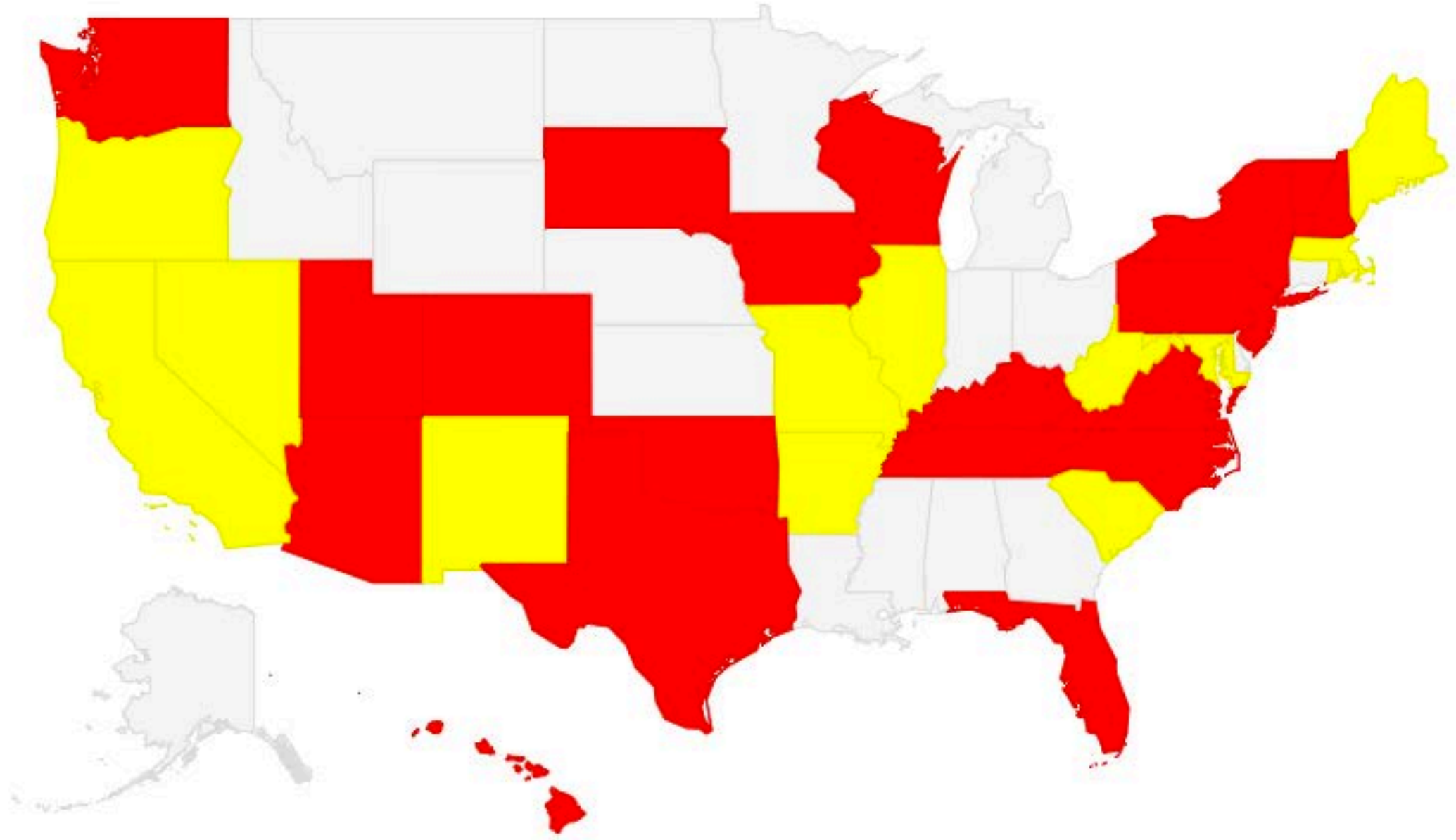

**Figure 2. United States map showing states where demographic data is HIPAA equivalent (yellow) or non-HIPAA equivalent (red). White states do not release data.**

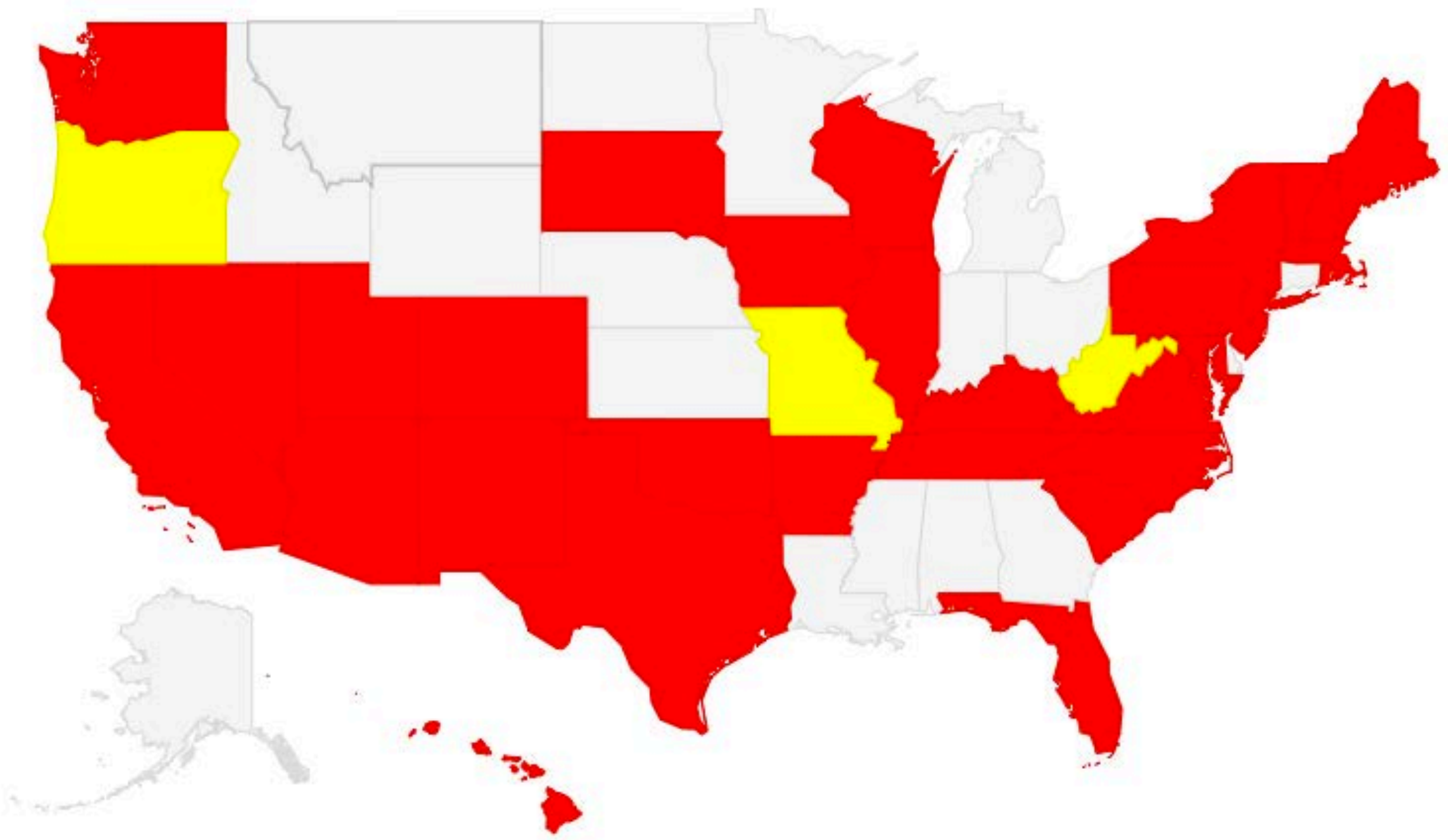

**Figure 3. United States map showing states where demographic *and* admission/discharge data is HIPAA equivalent (yellow) versus non-HIPAA equivalent (red). White states do not release data.**

## DISCUSSION

Is there vulnerability with using a standard less than HIPAA's? Is the HIPAA standard too stringent? Washington State releases data less strict than the HIPAA standard and in recent work, Sweeney showed how patients could be matched to records in the Washington State dataset to put names to the records [8]. Table 4 shows that Washington does not seem to be alone in its vulnerability to re-identification; re-identifications may be as possible on data from the other 30 states that release fields less than the HIPAA equivalent. If so, these vulnerabilities may threaten worthy and viable uses of the data unnecessarily.

Having more identifiable data readily available makes it difficult for other entities to share their data widely too. Data with some of the same fields as these hospital records becomes vulnerable to re-identification if the data to be shared can be linked to the more identifiable hospital data.

The goal is not to deprive society from the many worthy uses of the data made possible by sharing, but to match access requirements with risk, so society can enjoy the benefits of data sharing without unnecessary risks to patients. This seems achievable by making a public version of the data HIPAA equivalent and making more detailed information available under more stringent requirements.

Acknowledgments

The authors thank Amanda Black and Ryan Joyce for help locating and reviewing materials. The information presented here was culled from many sources and the best effort was made to collect the most accurate and up-to-date information. Please send the authors any updates or corrections for rectification. Check dataprivacylab.org/projects/50states and thedatamap.org for the latest information. Sean Hooley's work on this paper has been supported in part by a National Institutes of Health Grant (1R01ES021726), and Dr. Sweeney's in part by an NSF grant (CNS-1237235).